\documentclass[12pt]{iopart}
\pdfoutput=1

\usepackage[pdftex]{graphicx}
\usepackage{color}
\usepackage{mathptmx}
\usepackage{ulem}
\usepackage{bm}
\usepackage[pdftex,unicode,colorlinks]{hyperref}

\usepackage[dvipsnames]{xcolor}

\begin{document}

\title{Study of broadband multimode light via non-phase-matched sum frequency generation}
\author{Denis~A.~Kopylov$^{1,*}$ 
 Kirill~Yu.~Spasibko$^{2,3}$, Tatiana~V.~Murzina$^{1}$,  Maria~V.~Chekhova$^{2,3,1}$}
\address{
$^1$Department of Physics, M.V.~Lomonosov Moscow State University, Leninskie Gory GSP-1, 119991 Moscow, Russia
\\
$^2$ Max Planck Institute for the Science of Light, Staudtstr. 2, 91058 Erlangen, Germany
\\
$^3$ University of Erlangen-N\"urnberg, Staudtstr. 7/B2, 91058 Erlangen, Germany
}
\vspace{10pt}
\begin{indented}
\item[] October  2018
\end{indented}

\begin{abstract}
We propose non-phase-matched sum frequency generation (SFG) as a method for characterizing broadband multimode light. 
We demonstrate its advantages using high-gain parametric down conversion (PDC) as an example. 
By generating the sum frequency in the near-surface region of a nonlinear crystal, we increase the SFG efficiency and get rid of the modulation caused by chromatic dispersion, known as Maker fringes. 
The obtained results interpreted in terms of the Schmidt mode formalism reveal the sensitivity of the SFG spectrum to the number of frequency modes. 
Furthermore, we demonstrate efficient non-phase-matched three- and four-frequency summation of broadband multimode light, barely possible under phase matching.
\end{abstract}

\maketitle

\section{Introduction}

Sum frequency generation (SFG) is one of the most convenient tools for the study of spectral and temporal structure of light. 
Time-domain SFG is the key part of different pulse characterization techniques such as autocorrelation~\cite{Armstrong_1967}, FROG~\cite{Kane1993} or SPIDER~\cite{Iaconis1998}.
In particular, SFG is crucial for the study of nonclassical light~\cite{Pe_er_2005,O_Donnell_2009,Sensarn_2009,Jedrkiewicz_2012,Schwarz_2016}. 

For the SFG from broadband radiation, the well-known challenge is to make it efficient for the whole pump spectrum.
The standard solution to increase the bandwidth of SFG is a thin phase matched nonlinear crystal or a quasi-phase-matched crystal with aperiodic poling. 
However, the bandwidth of such devices is still limited; moreover, in the case of quasi-phase-matching SFG strongly depends on the quality of the nonlinear lattice.

In this paper we study non-phase-matched SFG in the bulk of a nonlinear crystal and near its surface. 
To compensate for the relatively low efficiency of non-phase-matched SFG, we use a crystal with high quadratic susceptibility (lithium niobate) in the geometry where its highest component is involved. 
By omitting the phase matching we drastically increase the frequency bandwidth and make the technique applicable to almost any input light. 
Furthermore, near the crystal surface we are able to observe broadband three- and four-frequency summation in the same geometry.

In our experiment, the input radiation is bright squeezed vacuum generated through the high-gain parametric down-conversion (PDC).
We demonstrate efficient broadband SFG on the surface of the crystal, revealing both coherent `pump reconstruction' and the incoherent background~\cite{Abram_1986,Jedrkiewicz_2011}. 
We describe these effects using Schmidt-mode formalism for PDC~\cite{Sharapova_2015, Pe_ina_2015, Christ_2013} and SFG with undepleted pump~\cite{Dayan_2007_theory}. 
For the SFG, an important issue is tight focusing of the pump beam~\cite{Sukhorukov_1980, Boyd_NO_2008}, which explains the difference between near-surface and bulk generation.

This paper is organized as follows. 
In Section~\ref{sec:theory} we describe the theory for SFG from tightly focused PDC. 
Section~\ref{sec:setup} presents the experimental setup and Section~\ref{sec:results}, the results. 
Section~\ref{sec:conclusion} concludes the paper. 

\section{Theory}\label{sec:theory}

\begin{figure*}[h]
\includegraphics[width=1.\linewidth]{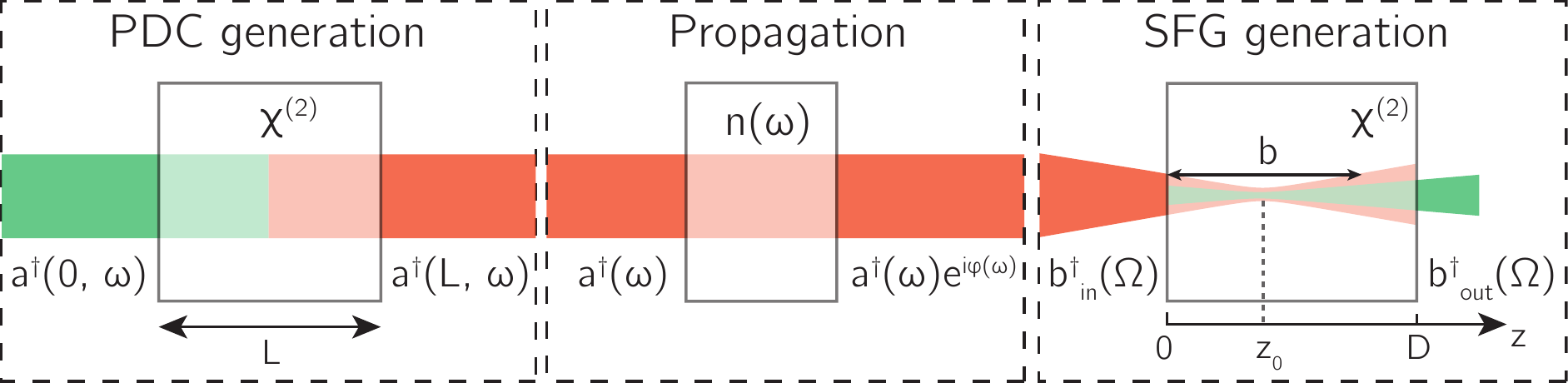}
\caption{Schematic representation of SFG from PDC.}
\label{fig_scheme}
\end{figure*}

In this section we present the theory of SFG from tightly focused broadband radiation. 
As a pump we use high-gain PDC radiation, also known as bright squeezed vacuum~\cite{Sharapova_2015}.
Figure~\ref{fig_scheme} shows schematically the three stages of the nonlinear interaction: generation of  broadband PDC radiation, its propagation in a dispersive medium, and SFG generation in a nonlinear crystal from a tightly focused beam. 
Further, we calculate the SFG spectrum by describing all three stages in a semi-analytical way.

\subsection{SFG from a tightly focused pump}

We describe the SFG using the theoretical approach of Ref.~\cite{Dayan_2007_theory}. 
In the undepleted pump approximation, the output plane-wave annihilation operator for SFG at frequency $\Omega$ is
\begin{equation}
\hat{b}_{out}(\Omega) = \hat{b}_{in}(\Omega) + \int \mathrm{d} \omega \;  K(\omega, \Omega) \,\hat{a}(\omega) \, \hat{a}(\Omega - \omega),
\label{init_SFG}
\end{equation}
where $\hat{b}_{in}$ is the input plane-wave annihilation operator. 
The transfer function $K(\omega, \Omega)$ for a plane-wave pump~\cite{Dayan_2007_theory},
\begin{eqnarray}
K_{pw}(\omega, \Omega) &=& \beta(\omega, \Omega) \int_0^D \mathrm{d} z \; \exp(-i \Delta \kappa(\omega, \Omega) z) \nonumber \\
&=& \beta(\omega, \Omega) \ D \ \mathrm{sinc} \left( \frac{\Delta \kappa(\omega, \Omega) D}{2} \right) \ \exp\left( \frac{ -i \Delta \kappa(\omega, \Omega) D}{2}\right),
\label{form_K_sinc}
\end{eqnarray}
where $\beta(\omega, \Omega)\sim \chi^{(2)} \sqrt{\omega(\Omega - \omega)\Omega}$, $\chi^{(2)}$ is the quadratic susceptibility, $\Delta \kappa(\omega, \Omega) \equiv \kappa(\Omega) - \kappa(\omega) - \kappa(\Omega - \omega) $ is the wavevector mismatch for SFG, and $D$ is the length of the SFG crystal.
For simplicity we assume $\beta(\omega, \Omega)\equiv\beta$ is a constant.

In the case of a tightly focused pump the plane-wave approximation does not hold.
In particular, the Gouy phase becomes important. 
To take it into account for SFG  from a Gaussian beam, we insert the factor $(1+2i(z-z_0)/b)^{-1}$ into the expression for the transfer function~\cite{Sukhorukov_1980,Boyd_NO_2008},
\begin{equation}
K_{G}(\omega, \Omega) = \beta \int_{0}^{D} \mathrm{d} z \; \frac{\exp(-i \Delta \kappa (\omega, \Omega) z)}{1+2i(z-z_0)/b}.  
\label{form_K_focus}
\end{equation}
Here, $b = \kappa_0 w_0^2$, $z_0$, and $w_0$ are the confocal parameter, the waist position and the radius of the pump beam, respectively, and $\kappa_0=n(\omega_0)\omega_0/c$ is the wavevector at the degenerate frequency $\omega_0$ (see Figure~\ref{fig_scheme}). 
For simplicity, we assume that the pump beam parameters are the same for all frequencies.
In the case of weak focusing, $b \gg D$, Eq.~(\ref{form_K_focus}) simplifies to Eq.~(\ref{form_K_sinc}). 

\subsection{PDC radiation}

In this paper the input radiation for SFG is multimode bright squeezed vacuum generated through high-gain PDC. 
Following the approach of Ref.~\cite{Pe_ina_2015}, we describe the PDC in terms of the averaged momentum operator. 
In the case of collinear PDC along the $z$ axis, it has the form
\begin{eqnarray}
\hat{G}_{int}(z) = \hbar \Gamma \int \int \mathrm{d} \omega_i \mathrm{d} \omega_s \; F(\omega_i,\omega_s) \hat{a}^\dagger(z, \omega_i) \hat{a}^\dagger(z, \omega_s) + h.c., 
\end{eqnarray}
where $\Gamma$ characterizes the interaction strength and $s, i$ indices denote the signal and idler waves, respectively. 
The joint spectral amplitude (JSA) $F(\omega_i,\omega_s)$ for the PDC generated in a nonlinear crystal with the length $L$ is \cite{Kim_2005},
\begin{eqnarray}
F(\omega_i,\omega_s) = f_p(\omega_i+\omega_s)
\textrm{sinc}\left(\frac{\Delta k(\omega_i,\omega_s) L}{2}\right) 
\exp\left( \frac{i\Delta k(\omega_i,\omega_s) L}{2} \right),
\end{eqnarray}  
where $f_p(\omega)$ is the pump spectral amplitude, $\Delta k(\omega_i,\omega_s) \equiv k_p(\omega_i + \omega_s) - k_{s}(\omega_s) - k_{i}(\omega_i) $ is the phase mismatch.
For the type-I phase matching, $k_p(\omega_p) = n_e(\omega_p, \theta)\omega_p/c$, $k_{s, i}(\omega_{s, i}) = n_o(\omega_{s, i})\omega_{s, i}/c$, where $n_o$ and $n_e$ are ordinary and extraordinary refractive indices and $\theta$ is the optic axis angle w.r.t. $k_p$.

The Heisenberg equation for the monochromatic-wave annihilation operators,
\begin{equation}
\frac{\partial \hat{a}(z, \omega)}{\partial z} = \frac{i}{\hbar}[\hat{a}(z, \omega), \hat{G}_{int}(z)],
\label{eq:heisenberg}
\end{equation}
can be analytically solved using the Schmidt decomposition of the JSA. 
The latter is symmetric for type-I phase matching, $F(\omega_i, \omega_s)=F(\omega_s, \omega_i)$, therefore
\begin{equation}
F(\omega_i, \omega_s) = \sum_n \sqrt{\lambda_n} \varphi_n(\omega_i) \varphi_n(\omega_s),
\end{equation}
where the Schmidt modes $\lbrace\varphi_n(\omega)\rbrace$ form an orthonormal basis and the Schmidt eigenvalues $\lambda_n$ are positive. 
Note that generally the JSA and $\varphi_n(\omega)$ are complex functions.

We introduce the Schmidt-mode creation and annihilation operators,
\begin{eqnarray}
\hat{A}_n^\dagger = \int \mathrm{d} \omega \; \varphi_n(\omega) \hat{a}^\dagger(0, \omega),\nonumber\\
\hat{A}_n         = \int \mathrm{d} \omega \; \varphi_n^{\ast}(\omega) \hat{a}(0, \omega).
\end{eqnarray} 
They have standard commutation relations, 
\begin{equation}
[\hat{A}_n, \hat{A}_m^\dagger]=\delta_{nm},\quad[\hat{A}_n, \hat{A}_m]=0.
\label{form_commutation_relations}
\end{equation}
Thus the solution of Eq.~(\ref{eq:heisenberg}) is \cite{Sharapova_2015}
\begin{equation}
\hat{a}(L, \omega) = \hat{a}(0, \omega)  + \sum_n \varphi_n(\omega) \Big( S_n \hat{A}^\dagger_n +(C_n - 1)\hat{A}_n \Big), 
\label{init_PDC}
\end{equation}
where $S_n = \sinh(\Gamma_n)$, $C_n = \cosh(\Gamma_n)$, and $\Gamma_n \equiv \Gamma\sqrt{\lambda_n}\,L$ is the parametric gain of the $n$-th mode.
Then the PDC spectral distribution of the photon number is 
\begin{equation}
N_{PDC}(\omega)=\langle 0| \hat{a}^\dagger(L, \omega)\hat{a}(L, \omega) |0\rangle  
= \sum_n S_n^2 |\varphi_n(\omega)|^2,
\label{from_pdc_spectrum}
\end{equation}
The effective number of Schmidt modes $K$ can be found as \cite{Sharapova_2015}
\begin{equation}
K = \Bigg[ \sum_n \Bigg( \frac{S_n^2}{\sum_m S_m^2}\Bigg)^2 \Bigg]^{-1}.
\label{eq:schmidt_weight}
\end{equation}
Here we find the Schmidt modes numerically using the Takagi factorization~\cite{Chebotarev_2014}, 
\begin{equation}
\mathcal{F} = \mathcal{U} \Sigma \mathcal{U}^T,
\label{Takagi}
\end{equation}
where $\mathcal{F}$ is a matrix corresponding to $F(\omega_i, \omega_s)$, $\mathcal{U}$ is a unitary matrix, and $\Sigma$ is a diagonal matrix with the singular values $\sigma_n = \sqrt{\lambda_n}$. 
The discretized Schmidt modes $\varphi_n(\omega)$ are given by columns of matrix $\mathcal{U}$.

\subsection{SFG from PDC}\label{sec:SFG_PDC}

According to Eq.~(\ref{init_SFG}), the SFG spectral distribution of the photon number is 
\begin{eqnarray}
N_{SFG}(\Omega)=\langle 0| \hat{b}_{out}^\dagger(\Omega)\hat{b}_{out}(\Omega) |0\rangle  
=\int \int \mathrm{d}\omega_1  \mathrm{d}\omega_2\nonumber\\ \times K^{\ast}(\omega_1, \Omega) K(\omega_2, \Omega)\langle 0|\hat{a}^\dagger(\omega_1)  \hat{a}^\dagger(\Omega - \omega_1) 
 \hat{a}(\omega_2)  \hat{a}(\Omega - \omega_2)|0\rangle .
\end{eqnarray}
We obtain
\begin{equation}
N_{SFG}(\Omega) =  \Big| \sum_n C_n S_n I_{nn}(\Omega) \Big|^2 
 + 2 \sum_{n, m} S^2_n  S^2_m \big|I_{nm}(\Omega)\big|^2,
\label{form_sfg_final}
\end{equation}

where $I_{nm}(\Omega)$ is the inter-modal interaction integral,
\begin{equation}
I_{nm}(\Omega) = \int \mathrm{d} \omega \; K(\omega, \Omega) \varphi_n(\omega) \varphi_m(\Omega -\omega). 
\label{form_integral_I}
\end{equation}
For a very thin crystal, $D \rightarrow 0$, $K(\omega, \Omega) \rightarrow const$; then $I(\Omega)_{nm}$ is proportional to the convolution of the Schmidt modes.

Expression~(\ref{form_sfg_final}) for the SFG spectrum contains two terms. 
The first one, so-called coherent contribution, corresponds to the SFG from each Schmidt mode with itself, i.e. the process reverse to PDC.
It results in a narrow spectral peak with the width determined by the PDC pump.
The second term, the incoherent contribution, is caused by the SFG from different (uncorrelated) Schmidt modes.
It gives a pedestal with the width determined by the whole PDC spectrum.

Previously the coherent and incoherent  SFG contributions were obtained for delta-correlated PDC~\cite{Dayan_2007_theory, Brambilla_2012}. 
Our result~(\ref{form_sfg_final}) takes into account the finite width of photon correlations, which is significant for PDC generated by short pulses.
In particular, it shows that the width of the `reconstructed' peak  is not the same as the one of the pump; it is somewhat broader. 

Note that here we considered the case where PDC has a single spatial mode. 
This approach is valid for waveguide PDC sources \cite{Eckstein_2011},  for specially engineered PDC sources~\cite{P_rez_2014} and in the presence of spatial filtering (as in our experiment). 
In the general case, PDC is multi-mode both in time and in space, and one should take into account the overlap between PDC and SFG spatial modes.

\subsection{Dispersion effects in SFG}\label{sec:dispersion}
In the course of propagation in a dielectric medium, a broadband radiation undergoes dispersion spreading.
We take it into account through the replacement $\hat{a}^\dagger(L, \omega) \rightarrow \hat{a}^\dagger(L, \omega)e^{i\phi(\omega)}$, where $\phi(\omega)$ is the PDC nonlinear phase shift (chirp). 
This substitution leads to the replacement $\varphi_n(\omega) \rightarrow \varphi_n(\omega)e^{i\phi(\omega)}$ and $\varphi_m(\Omega -\omega) \rightarrow \varphi_m(\Omega -\omega)e^{i\phi(\Omega -\omega)}$ in Eq.~(\ref{form_integral_I}).

\section{Experimental setup}\label{sec:setup}
\begin{figure*}
\includegraphics[width=1.\textwidth]{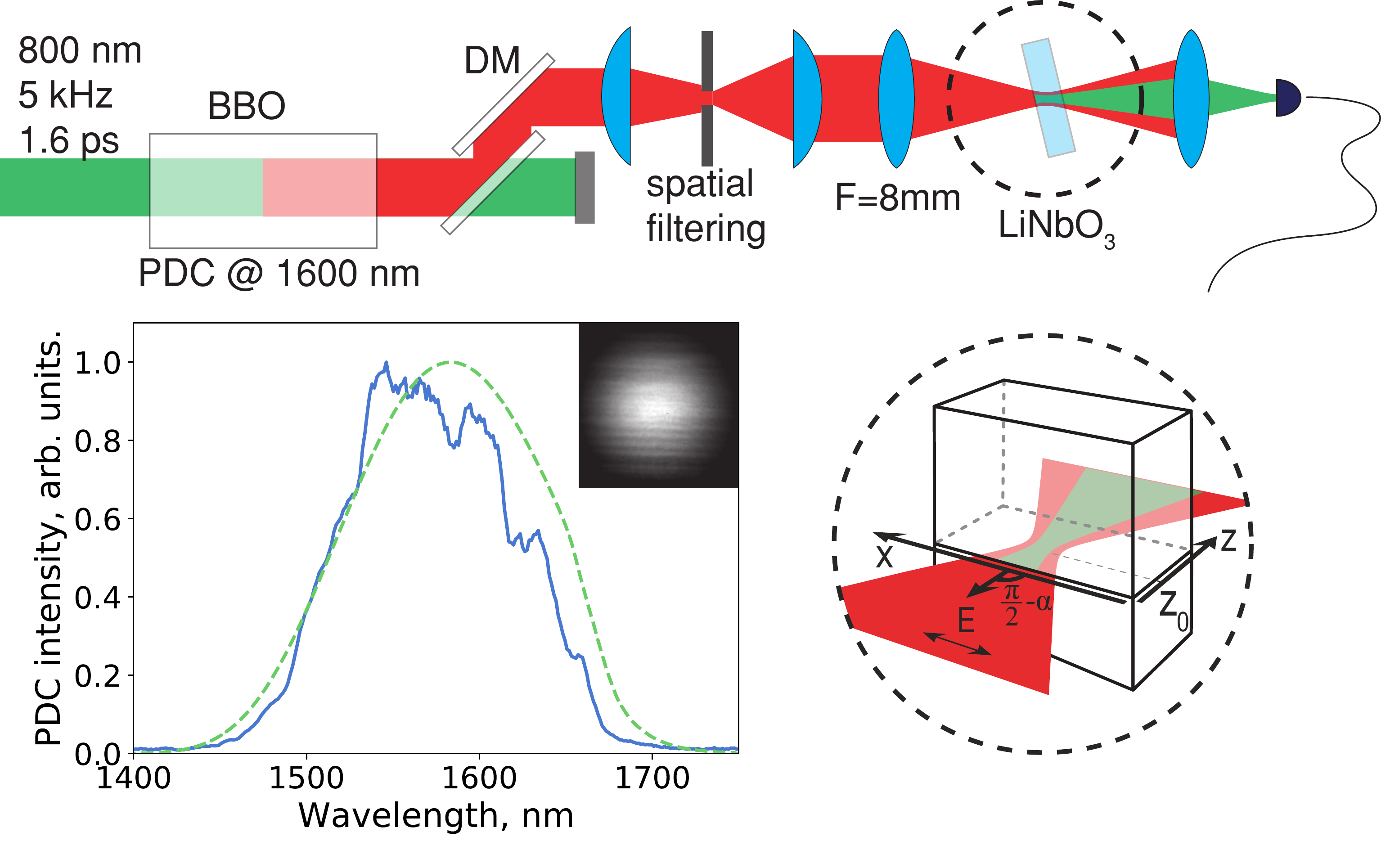}
\caption{ Top: experimental setup. Bottom left: measured (blue solid line) and calculated (green dashed line) PDC spectra and PDC beam after spatial filtering (inset). Bottom right: sketch of tight focusing SFG.}
\label{fig_setup}
\end{figure*}

Our experimental setup is shown in Figure~\ref{fig_setup}~(top). 
PDC is produced in a 10~mm BBO crystal with type-I collinear and frequency degenerate phase matching \cite{Spasibko_2016}.
The pump, propagating at $\sim$19.9$^{\circ}$  w.r.t. the optic axis, is the radiation of an amplified Ti:Sapphire laser. It has the central wavelength 800~nm, the pulse duration 1.6~ps, the repetition rate 5~kHz and the mean power 3~W.
To reduce the directional amplification caused by the spatial walk-off \cite{P_rez_2015}, the pump is focused into the crystal by means of a cylindrical lens with the 700~mm focal length.
The pump remains unfocused and relatively big ($2.6$~mm) in the horizontal direction.
It is cut off by two dielectric dichroic mirrors. 

Due to pump focusing, PDC is single-mode in the vertical direction. In the horizontal one we filter it to a single mode
with 140~$\mu$m vertical slit in a focal plane of 200~mm lens. 
The beam is collimated by a 300~mm cylindrical lens thereafter; a Gaussian beam with an extremely broadband spectrum (FWHM of 140~nm) is formed.
Figure~\ref{fig_setup}~(bottom left) shows the beam (inset) and the spectrum, both calculated (green line) and measured with an IR spectrometer (Hamamatsu-C11482GA, blue line).
We measure the parametric gain $\Gamma_0=10.5$ for the first Schmidt mode.
The difference between experimental and numerical spectra is introduced by dielectric mirrors that are used in the experimental setup (not shown in Fig.~\ref{fig_setup}). 

The PDC beam is focused into a 1~mm LiNbO$_3$ crystal doped with 5.1\% of Mg~\cite{Kitaeva_2000, Spasibko_2017} by means of an aspherical lens with the $8$~mm focal length. 
Such focusing results in the beam waist $2w_0 \approx 5~\mu$m and the confocal parameter $b\approx 40~\mu$m.
The LiNbO$_3$ crystal has its optical axis parallel to the facet, therefore the SFG occurs through the $ee \rightarrow e$ interaction.
For LiNbO$_3$ this interaction has very large quadratic susceptibility component $\chi^{(2)} = 2 d_{33}\sim60$~pm/V.
In our measurements we tune the angle of incidence $\alpha$  and beam waist position $z_0$ w.r.t the crystal (Figure~\ref{fig_setup}, bottom right).
The resulting SFG spectrum is measured with a visible spectrometer (Avantes AvaSpec-ULS3648).

\section{Results and discussion}\label{sec:results}
\begin{figure*}
\includegraphics[width=0.5\textwidth]{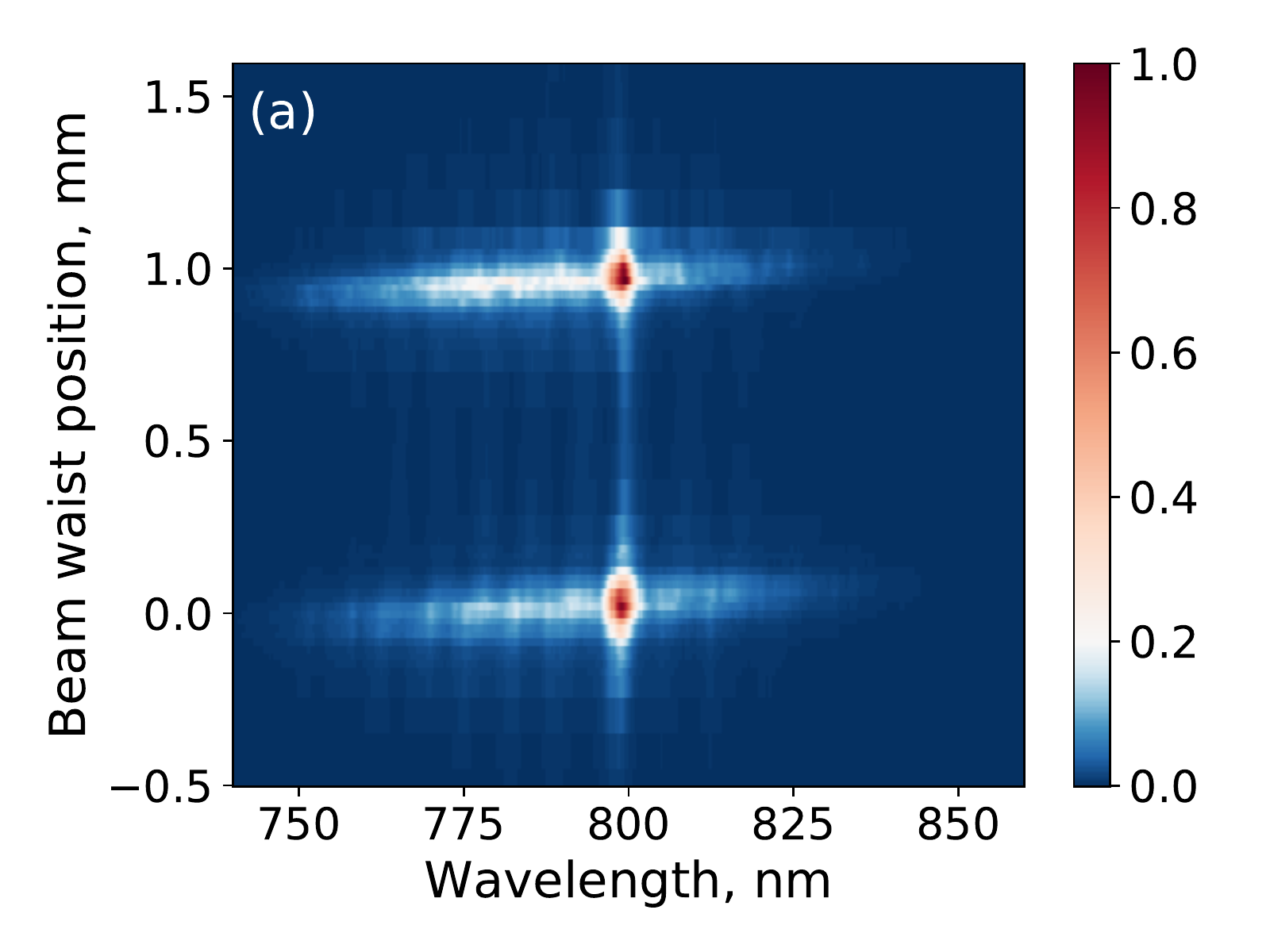}
\includegraphics[width=0.5\textwidth]{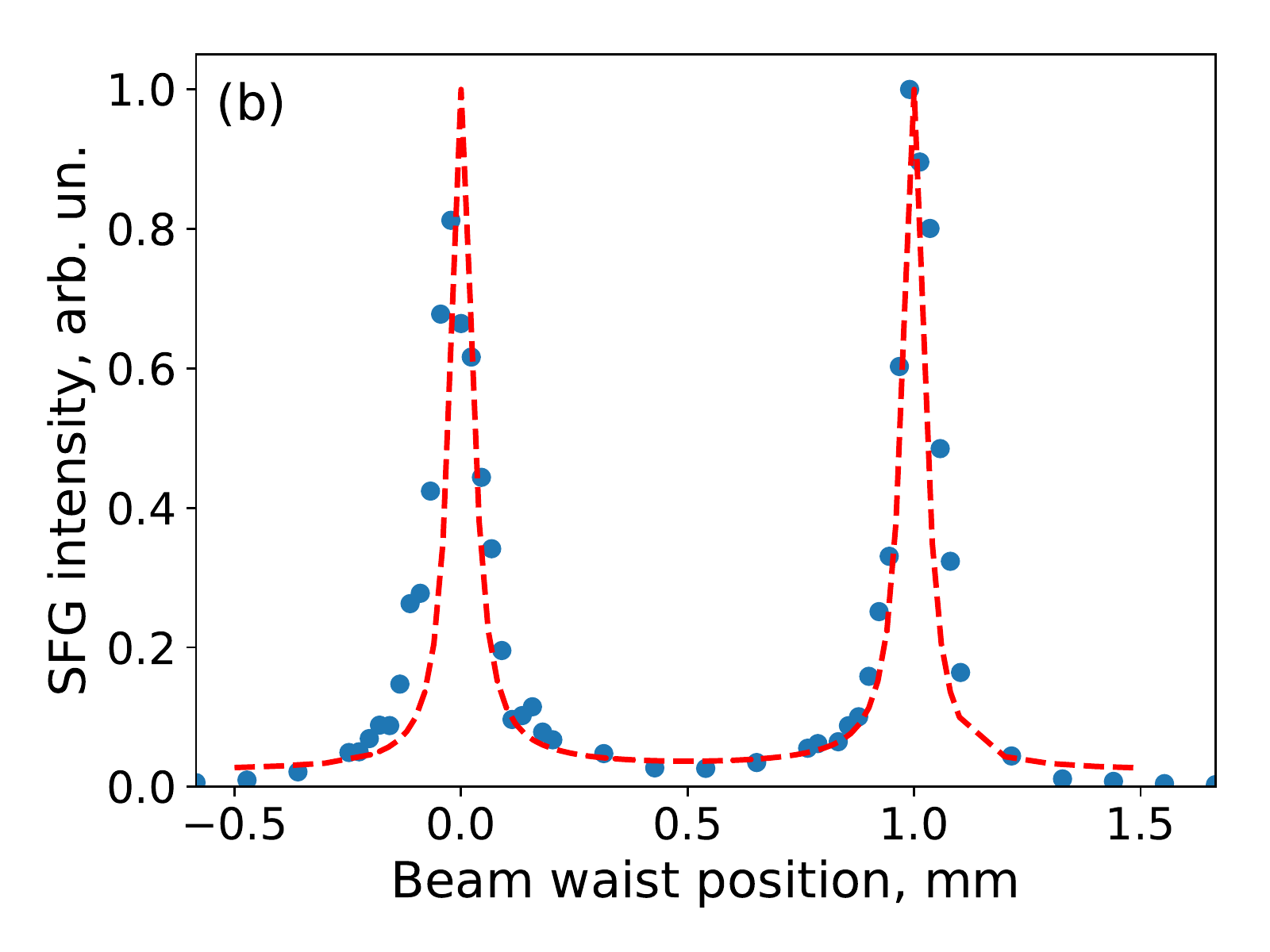}
\caption{SFG spectrum (a) and SFG signal at $800$~nm (b), measured (blue dots) and calculated (red line), versus the PDC beam waist position.}
\label{fig_2d_spatial}
\end{figure*}

Figure~\ref{fig_2d_spatial}a shows how the SFG spectrum depends on the beam waist position inside the crystal.
The SFG intensity takes its highest values when PDC is focused on one of the crystal facets; focusing in the bulk leads to inefficient SFG.

Such behavior is caused by the Gouy phase for the pump radiation and has nothing to do with so-called surface nonlinear susceptibility.
The Gouy phase differs by $\pi$ before and after the waist; therefore it leads to an additional phase shift in the nonlinear polarization. 
As a result, the SFG contributions before and after the waist interfere destructively, which leads to a low SFG intensity if the waist is in the bulk. 
Variation of the beam focusing or the crystal length strongly affects this behavior.

The theory from Eq.~(\ref{form_K_focus}, \ref{form_sfg_final}, \ref{form_integral_I}) predicts the same effect.
Theoretical SFG signal at $800$~nm (Figure~\ref{fig_2d_spatial}b, red line) agrees with the experimental one (blue dots).
The latter is a 1D cross-section of Figure~\ref{fig_2d_spatial}a.

In our setup we are not able to measure the group delay dispersion (GDD). Therefore we use it as a fitting parameter (see Section~\ref{sec:dispersion}).
The best agreement between the theory and experiment is obtained for GDD $\approx 200$~fs$^2$.
Furthermore, in order to take into account that the detection happens in wavelength-angular space, not in frequency-wavevector one, the calculated SFG spectra are corrected by the factor $\lambda^{-4}$ in accordance with~\cite{klyshko1988book, Spasibko_2012}.

\begin{figure*}
\includegraphics[width=0.5\textwidth]{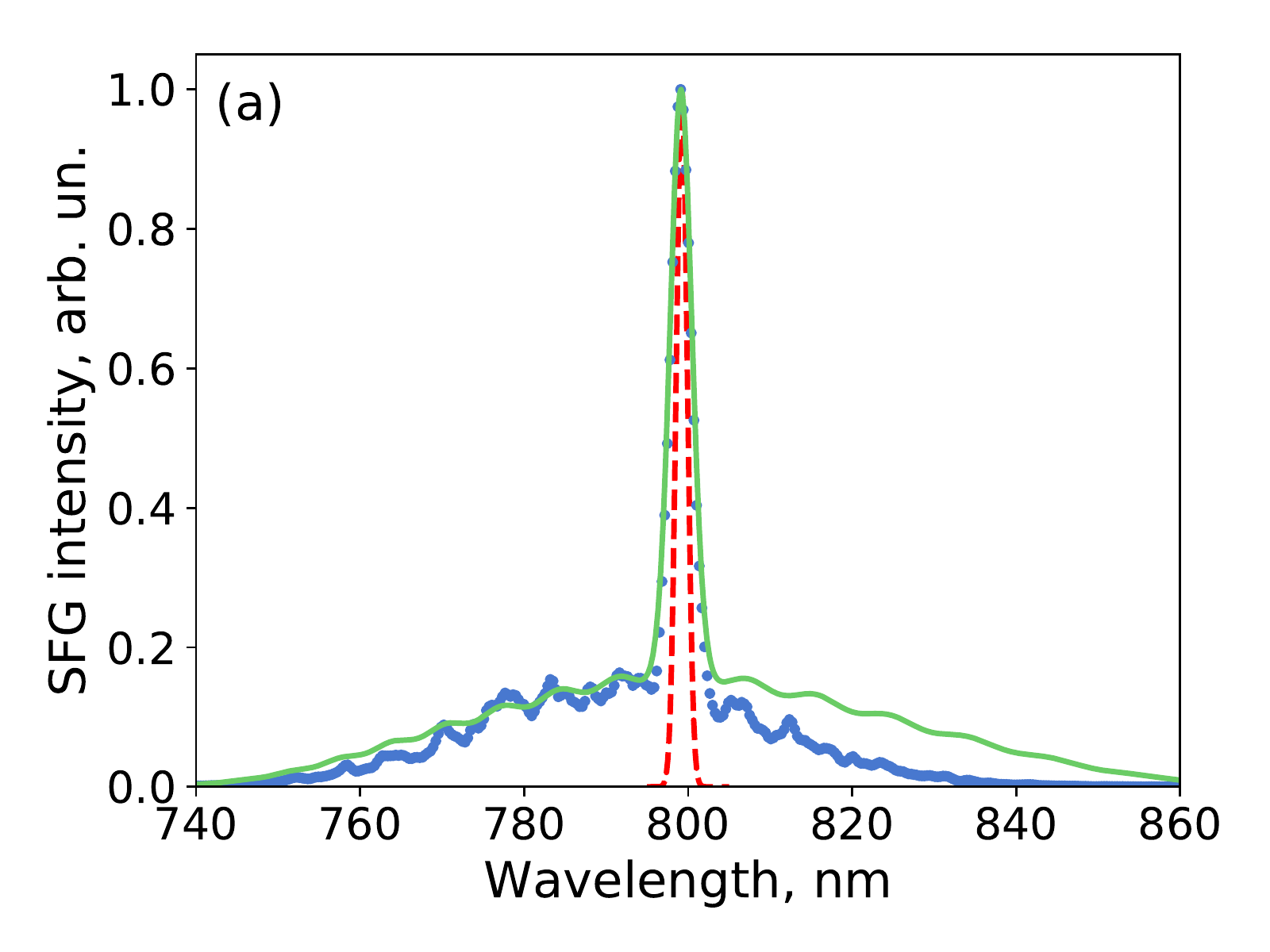}
\includegraphics[width=0.5\textwidth]{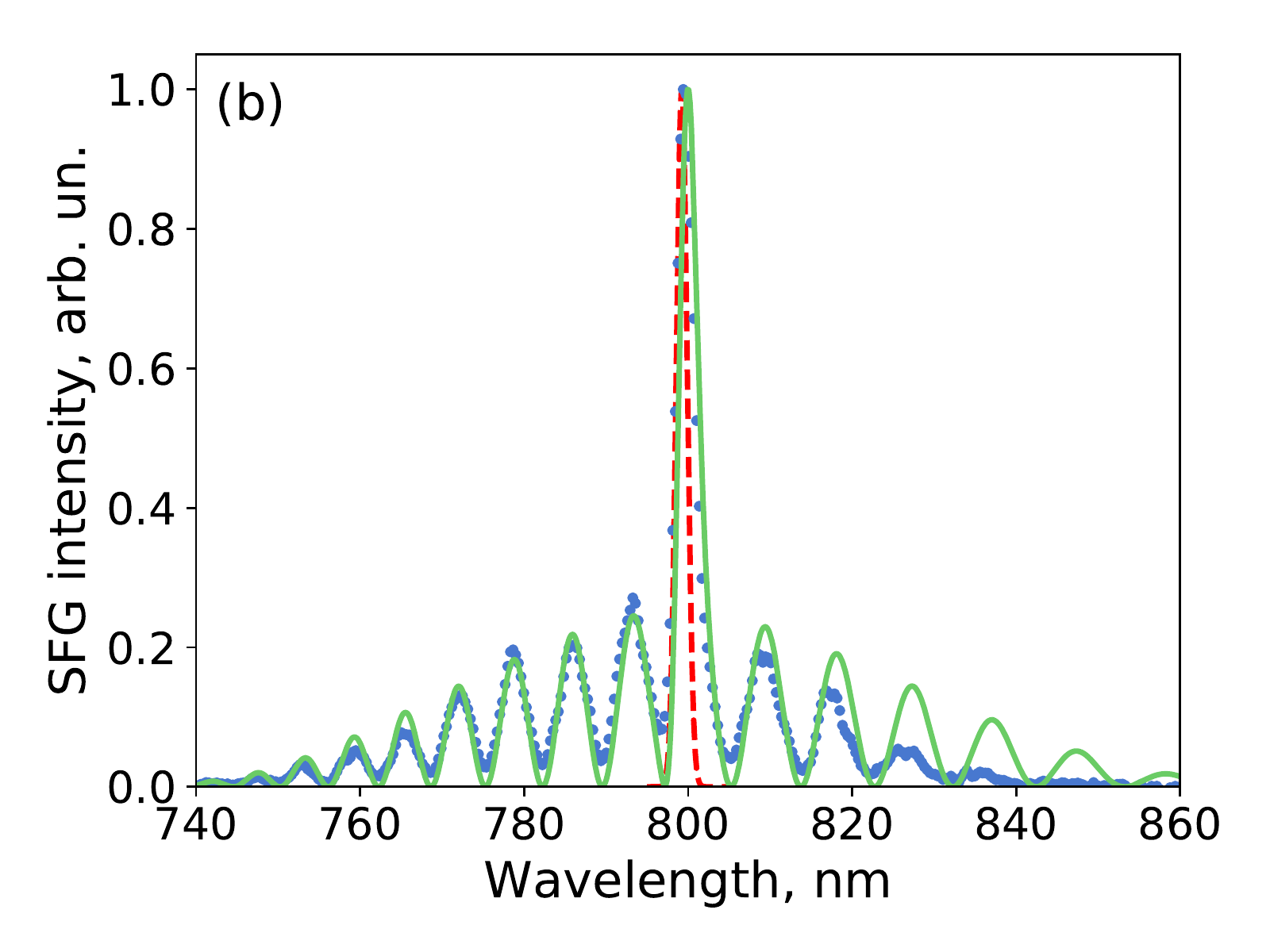}
\caption{
The experimental (blue dots) and theoretical (green line) SFG spectra for PDC focused (a) onto the facet and (b) into the bulk of the LiNbO$_3$ crystal.
Red dashed line shows the spectrum of the Ti:Sapphire laser.}
\label{fig_center_and_edge}
\end{figure*}

\begin{figure*}
\includegraphics[width=0.5\textwidth]{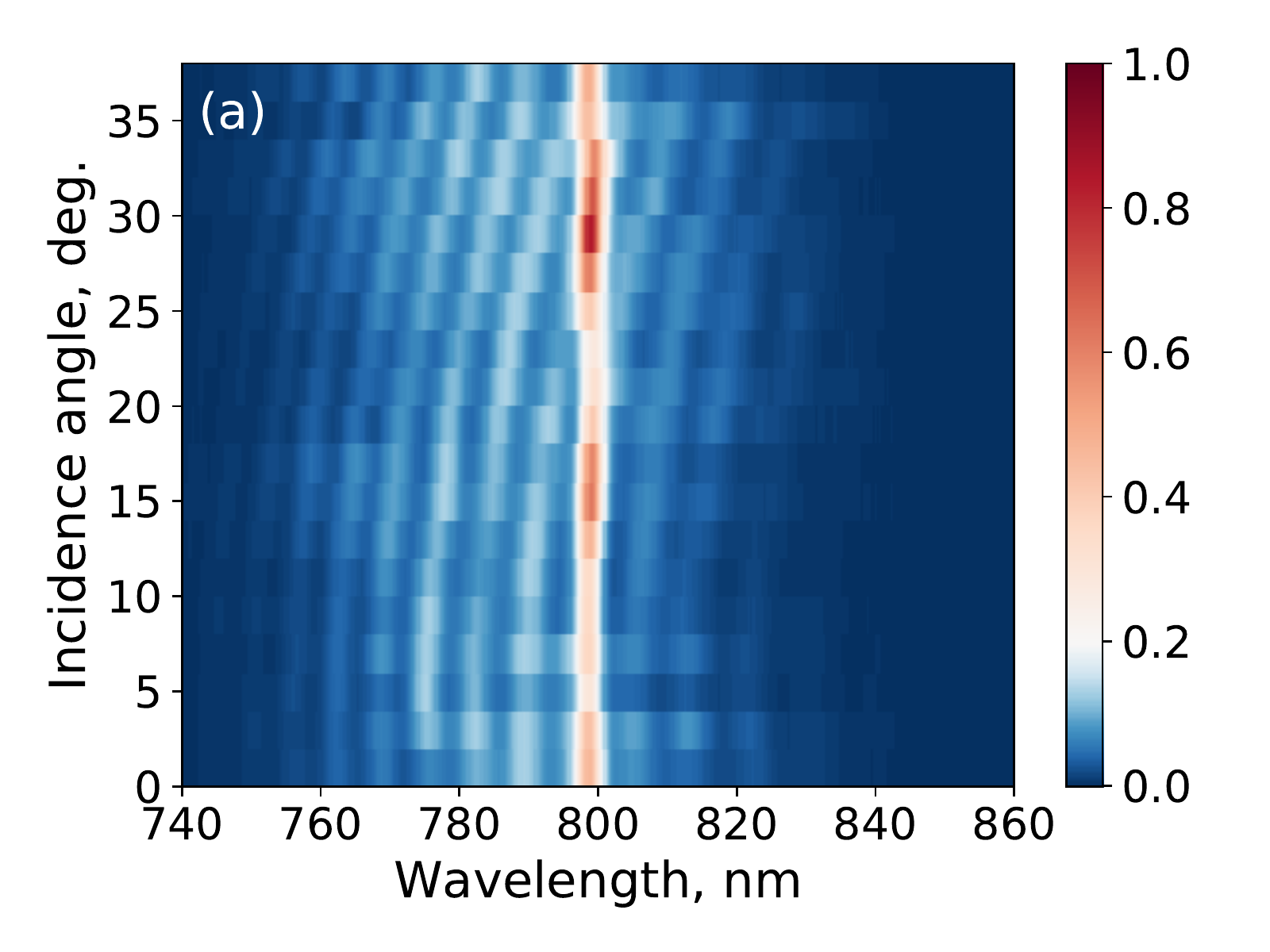}
\includegraphics[width=0.5\textwidth]{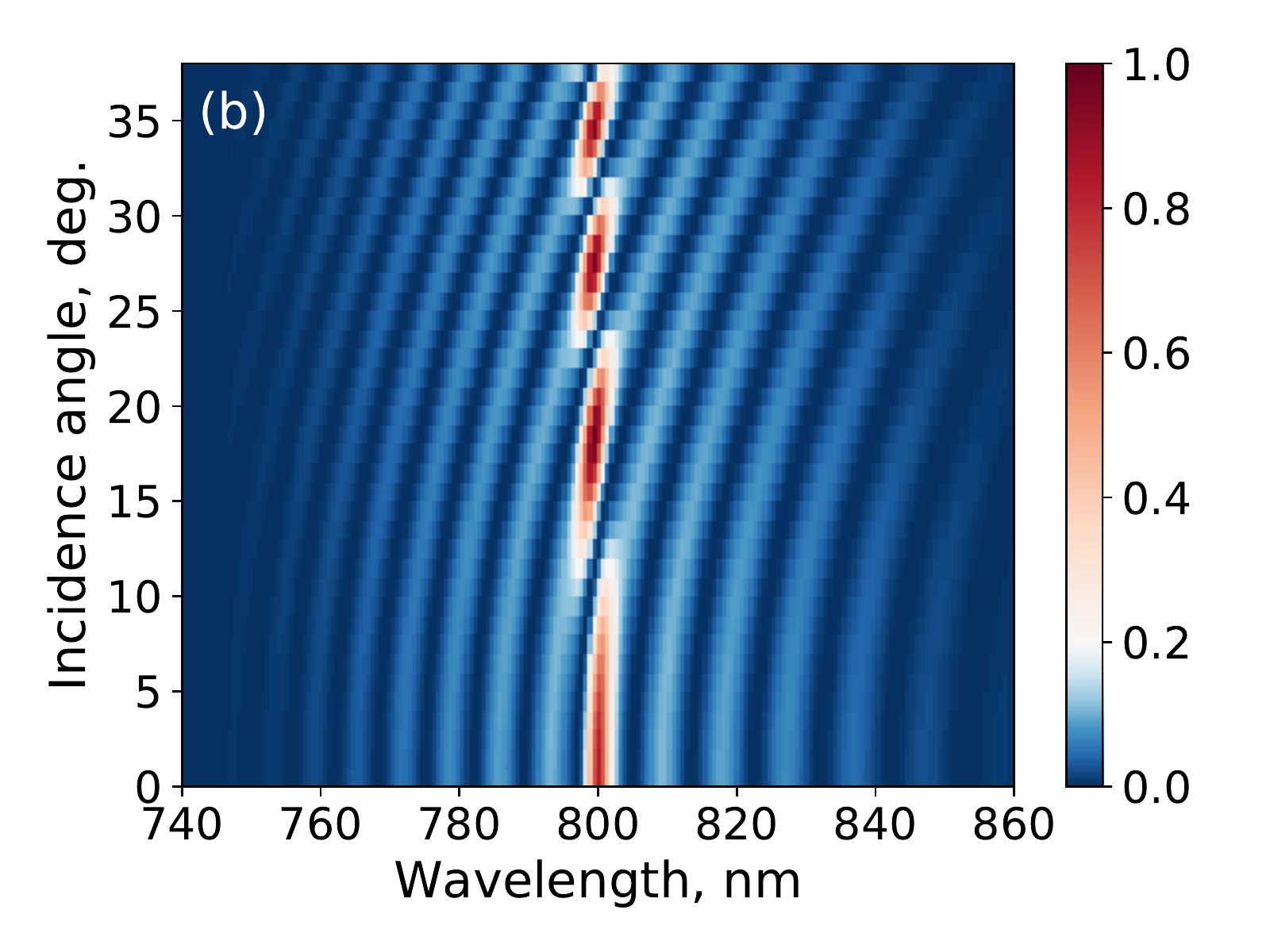}
\caption{The experimental (a) and theoretical (b) SFG spectrum versus the angle of incidence $\alpha$ for the bulk focusing.}
\label{fig_angular}
\end{figure*}

Both theoretical (blue dots) and experimental (green line) SFG spectra are shown in Figure~\ref{fig_center_and_edge}.
Panel (a) shows the case of focusing onto the facet ($z_0=0$~mm) and panel (b), into the bulk ($z_0=0.5$~mm).
The spectrum strongly depends on the focusing.
In the bulk case, high-visibility fringes, similar to the Maker ones~\cite{Maker_1962}, appear due to the variation of the wavevector mismatch $\Delta\kappa$ with the wavelength.
Their spectral positions depend on the angle of incidence  $\alpha$ (Figure~\ref{fig_angular}); the theory predicts similar behavior.
The fringe period scales as the inverse length of the crystal.

As discussed in Section~\ref{sec:SFG_PDC}, the spectrum contains two parts, the peak and the background, and the peak is usually interpreted as the pump reconstruction~\cite{Abram_1986,Jedrkiewicz_2011}.
However, for high-gain PDC the peak is broader than the pump spectrum (red dashed line), both in theory and in the experiment.
This broadening comes from the decrease of the number of modes with the parametric gain~\cite{Sharapova_2015}.
The background is narrower than predicted by the theory due to the difference between the real PDC spectrum and the theoretical one (Figure~\ref{fig_setup}b).

SFG can be interpreted as an ultrafast correlator or coincidence circuit~\cite{Sensarn_2010}.
Expanding this analogy further, one can set the correspondence between the SFG spectrum and the spectral distribution of coincidences used to measure the squared modulus of the JSA, the joint spectral intensity (JSI)~\cite{Di_Lorenzo_Pires_2009}. 
Then, the width of the peak $\Delta \omega_{coh}$ corresponds to the `conditional width'  of the JSI, which is the spectral distribution of the coincidence rate for a fixed frequency of the idler photon. 
Meanwhile, the width of the background $\Delta \omega_{incoh}$ corresponds to the JSI `unconditional width', which is the spectral width of the signal radiation. 
The ratio of the unconditional and conditional widths, according to Ref.~\cite{Fedorov_2006}, can be used to assess the degree of frequency entanglement for photon pairs, as it is close to the number of the Schmidt modes in the PDC spectrum. 
It follows that the ratio of the spectral widths, $R = \Delta \omega_{incoh}/\Delta \omega_{coh}$, can be also interpreted as the Schmidt number~$K$.

\begin{figure*}
\begin{center}
\includegraphics[width=0.5\textwidth]{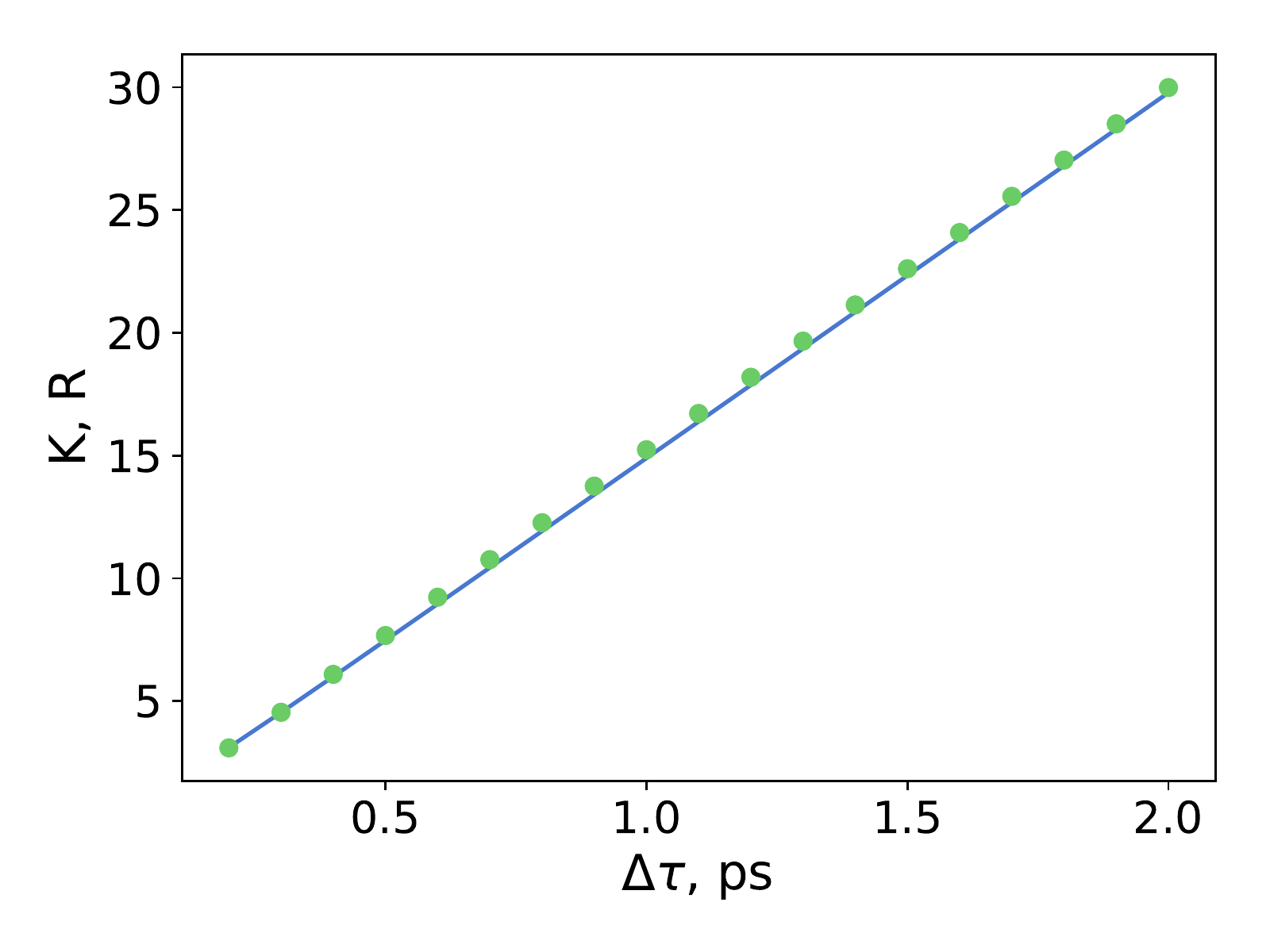}
\end{center}
\caption{
The Schmidt number $K$ (blue line) and the ratio $R$ (green dots) calculated as a function of the pump pulse duration.}
\label{fig:ratio}
\end{figure*}

To test this statement, we have calculated both the Schmidt number $K$, using Eq.(\ref{eq:schmidt_weight}), and the resulting ratio $R$ for different values of the pump pulse duration. The result is shown in Fig.~\ref{fig:ratio}; as expected, we see that $R=K$.
In experiment we get $R_{exp} = 12.7$, which is smaller than the expected one $K_{theor} = R_{theor} = 18$, because the theoretical SFG spectrum is different from the experimental one.

Furthermore, non-phase-matched generation is also possible for higher-order nonlinear optical effects where it is usually difficult to satisfy the phase matching.
Similarly to the SFG case considered above, strong focusing in the bulk can completely suppress high-order nonlinear optical processes~\cite{Sukhorukov_1980, Boyd_NO_2008}. 
Focusing near the surface allows one to increase their efficiency as well.

\begin{figure*}
\includegraphics[width=0.5\textwidth]{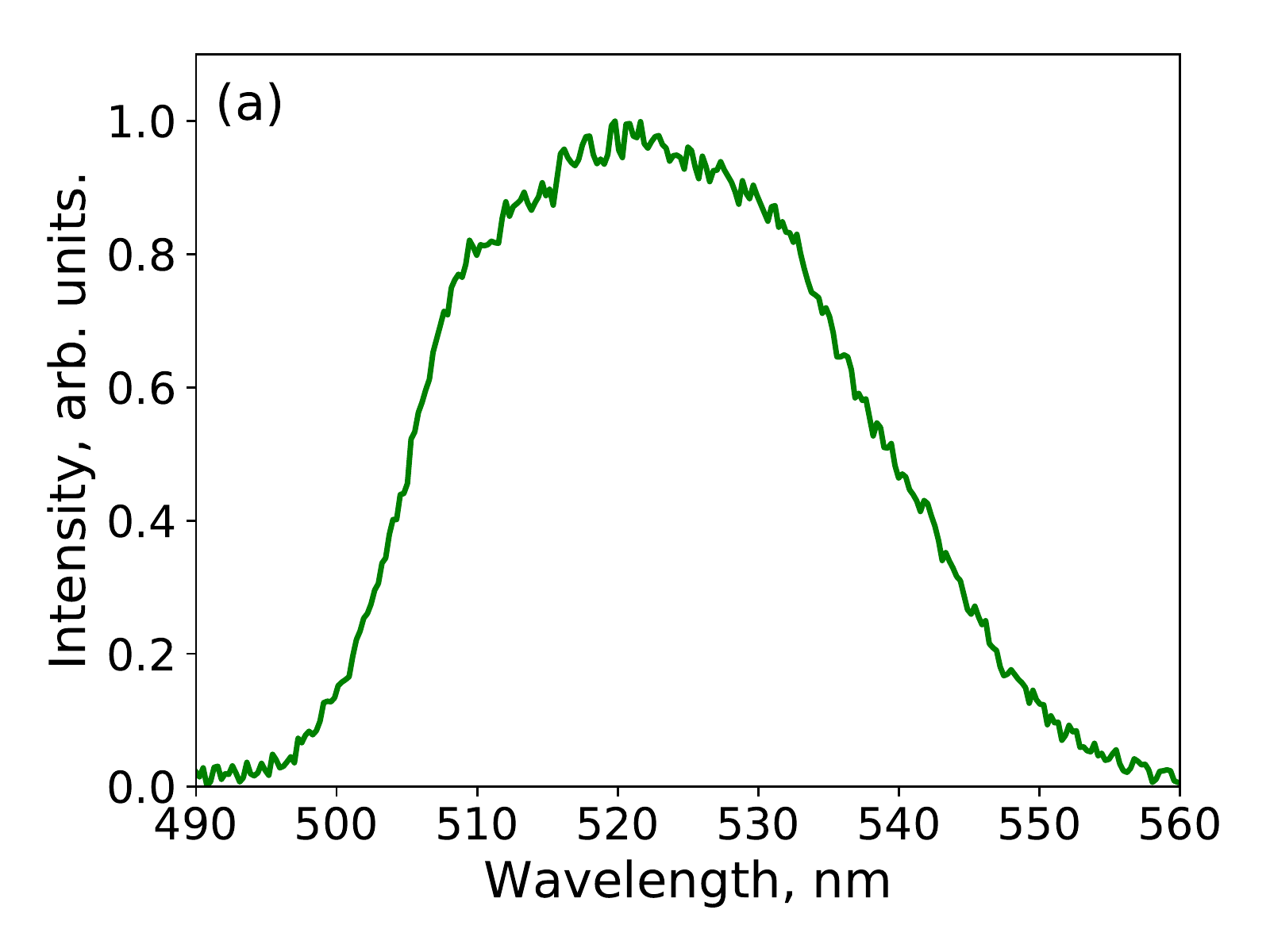}
\includegraphics[width=0.5\textwidth]{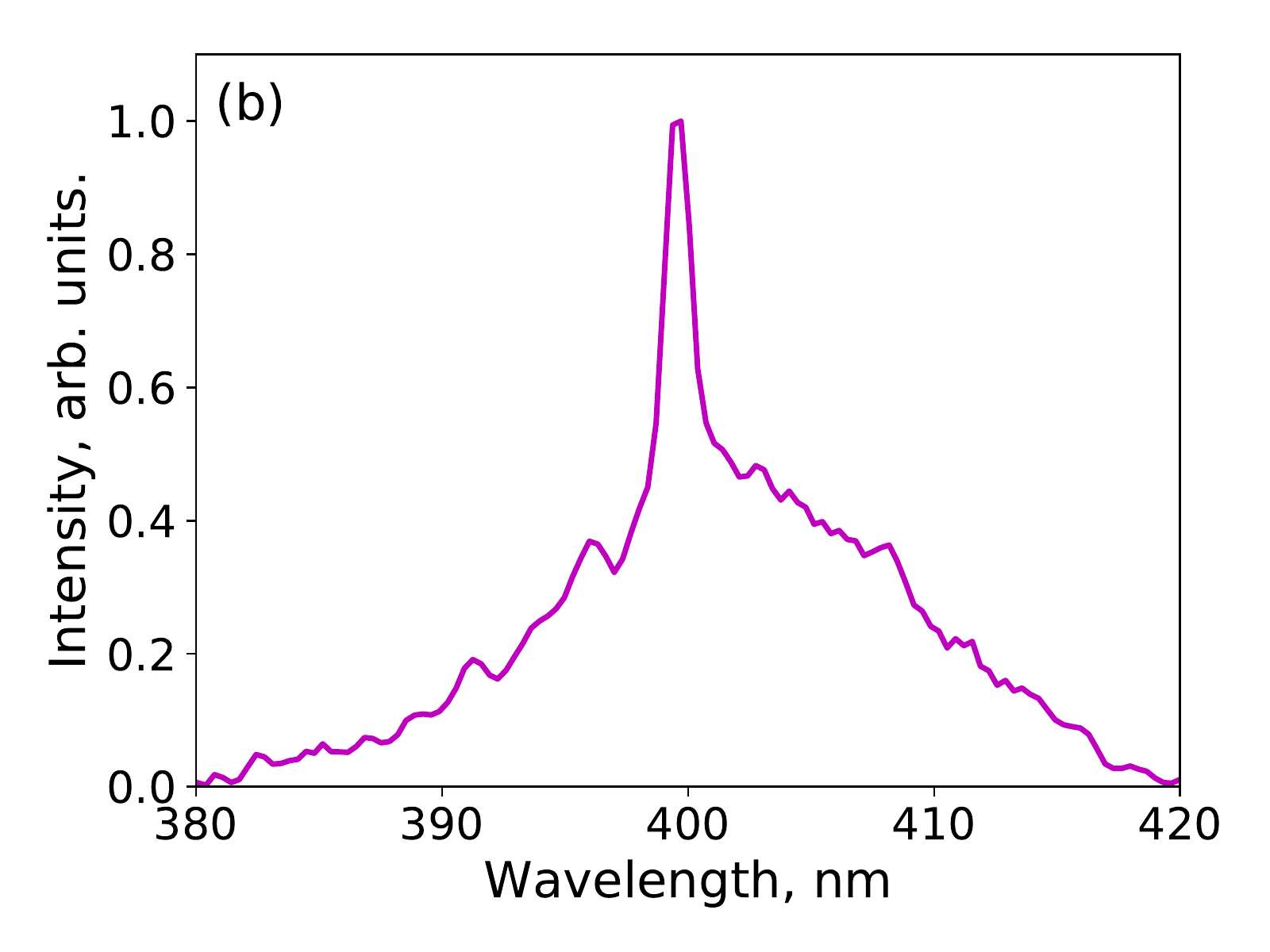}
\caption{ The three-frequency summation (TFS, left) and four-frequency summation (FFS, right) generated in the LiNbO$_3$. The PDC is focused in the facet of the crystal.}
\label{fig_third_fourth}
\end{figure*}

To demonstrate it, we observe the three- and four-frequency summation of multimode PDC radiation near the surface of the same LiNbO$_3$ crystal.
Figure~\ref{fig_third_fourth} shows the measured spectra.
Similarly to the case of SFG, for four-frequency summation we observe the coherent peak and the incoherent background, while for three-frequency summation there is no peak.
This behavior, the even harmonics containing a peak and the odd ones, not, has been predicted theoretically for two-photon correlated light~\cite{Masalov1991} but never observed in experiment, to the best of our knowledge. 
Qualitatively, this effect can be understood by considering that photons are produced in pairs: the pump is reconstructed only if both photons of a pair are involved into the process.
A rigorous theoretical consideration can be found in Ref.~\cite{Masalov1991}.

Previously, the coherent and incoherent contributions of PDC were observed only for SFG and only under phase matching~\cite{Jedrkiewicz_2012,Abram_1986,Jedrkiewicz_2011,  quant-ph/0302038}.
Our results show that even without phase matching a pronounced narrowband coherent contribution and a broad incoherent contribution can be observed. 
Moreover, focusing near the surface of a nonlinear crystal suppresses the Maker fringes, which offers a possibility to use SFG as not just a broadband, but a \textit{wavelength-independent} autocorrelator~\cite{Chekhova_2018}. This type of autocorrelator will have an advantage over the one based on two-photon absorption (TPA) in semiconductors~\cite{Takagi_1992,Boitier_2011}. Indeed, the operating wavelength of a TPA autocorrelator is limited for the wavelengths that satisfy the two-photon transition between the valence and conduction energy zones of a semiconductor band-gap.
In the case of tightly focused non-phase matched SFG in the near-surface domain, the operating wavelength is only limited by the transparency window of the nonlinear crystal. 
Moreover, even in the presence of  absorption at the sum frequency it is possible to measure the temporal characteristics of light~\cite{Kintzer_1987}.

\section{Conclusion}\label{sec:conclusion}

We have shown that non-phase matched SFG can be used to study the spectral properties of broadband light. In particular, with the radiation of high-gain PDC at the input, the SFG spectra show both the coherent contribution (a narrow peak) and the incoherent contribution (a broad background), which so far have been only observed for phase matched or quasi-phase matched SFG.
By tightly focusing the radiation on the surface of the crystal instead of the bulk, one can strongly increase the efficiency of SFG; otherwise the Gouy phase leads to the destructive nonlinear interference.
Moreover, by using the surface one gets rid of the Maker fringes, which modulate the spectrum in the case of bulk SFG.

The obtained experimental results allow a simple interpretation within the Schmidt-mode formalism. In particular, we have found that the Schmidt number is equal, up to a constant close to unity, to the ratio of the widths of the incoherent background and the coherent peak observed in the SFG spectra. 

The same geometry, with tight focusing of the input PDC radiation on the surface of the same crystal, allowed us to observe, in addition to SFG, the three- and four-frequency summation. As expected from the theory, for the
four-frequency summation, the spectrum consists of a pronounced coherent peak and incoherent background, while the three-frequency summation does not manifest the peak.

We believe that under tight focusing, non-phase matched SFG and its higher-order analogues can be used for the characterisation of nonclassical light sources and form the base for an ultrafast wavelength-independent autocorrelator.

\section{Acknowledgments}

We thank R. Penjweini for helpful discussions. M.V.C and K.Yu.S acknowledge the financial support of the joint DFG--RFBR (Deutsche Forschungsgemeinschaft -- Russian Foundation for Basic Research) Project No.~CH1591/2-1 -- 16-52-12031~NNIOa.
D.A.K acknowledge the financial support of Russian Foundation for Basic Research Project No.~18-32-00710 and German-Russian Interdisciplinary Science Center Project No.~P-2016a-5, P-2017a-19, P-2018a-16.

\section*{References}

\bibliographystyle{unsrt}
\bibliography{maker_lit}

\end{document}